\documentclass{article}

\usepackage{graphicx}
\usepackage[round]{natbib}
\usepackage{psfig}
\usepackage{epsfig}

\title{
Comparing classical and Bayesian methods for predicting hurricane landfall rates
 }

\author{Tim Hall, GISS\\and\\
Stephen Jewson\footnote{\emph{Correspondence address}: Email:
\texttt{x@stephenjewson.com}}\\}

\begin{document}
\maketitle

\begin{abstract}
We compare classical and Bayesian methods for fitting the poisson distribution to the number
of hurricanes making landfall on sections of the US coastline.
\end{abstract}

%

\section{Introduction}

The occurrence of a number of hurricanes making landfall on the US coastline in 2004 and
2005 has increased the level of interest in the question of how to estimate the risk
of hurricane damage in different locations. There are many aspects to this problem,
such as questions about the number of hurricanes occurring in different regions, their size,
their intensity, their speed, the detailed structure of their windfield, how they decay when they
reach land, how much damage the winds will cause to structures, and so on. In this paper,
we address one small part of one of these questions: how to estimate the distribution
of the number of hurricanes that make landfall on any short stretch of coastline using historical hurricane
data. This question itself has various aspects to it, such as to what extent we can extrapolate
historical hurricane data in space and to what extent the changing climate is changing the distribution
of the number of hurricanes. We don't attempt to deal with all these issues (at least not
here and now). Rather, we focus on one very small part of the problem, which is to look at
the most simple statistical methods that one might use to estimate hurricane landfall rates, ignoring the
complex issue of climate change completely. Given the historical data on hurricane landfalls,
we consider three basic methods one can consider using to estimate the distribution of hurricanes hitting a
certain region: the first is the classical method of moments,
the second is the classical maximum likelihood fitting procedure, and the third is
a Bayesian prediction procedure. The point of this paper is to discuss these three procedures, and
compare their pros and cons when applied to this particular problem.

In section~\ref{setup} we describe the question we want to address in more detail.
In section~\ref{classical} we describe the classical methods.
In section~\ref{bayesian} we describe the Bayesian method.
In section~\ref{numerical} we perform a numerical comparison of the two methods.
In section~\ref{scoring} we discuss one theoretical way in which these methods can be compared.
In section~\ref{empirical} we apply the two methods to observed hurricane landfall data,
and compare the results using cross-validation.
Finally in section~\ref{discuss} we discuss what we find.

\section{Setup}\label{setup}

We now give a mathematical summary of the problem we are considering.

We consider a short section of the US coastline, and we wish to estimate
the distribution of the number of hurricanes crossing this section of coastline.
We have $m$ years of historical hurricane data that we can use to estimate this distribution.
In practice $m$ has values of between 50 and 150 years, depending on the extent to which one
is willing to use earlier and hence less accurate data. In our examples below we will use the 54 years
of data from 1950 to 2003.

\begin{figure}[!hb]
  \begin{center}
    \scalebox{0.7}{\includegraphics{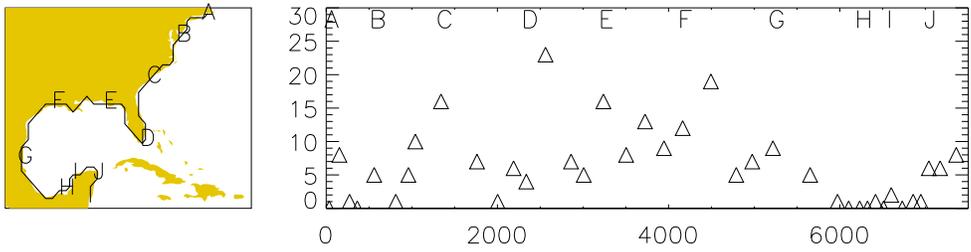}}
  \end{center}
    \caption{
The observed numbers of hurricanes crossing 39 straight-line segments
approximating the North American coastline, for the period 1950-2003.     }
     \label{f00}
\end{figure}

During these $m$ years of history there have been $i$ hurricanes making landfall on our section of coastline.
To illustrate possible values of $i$, figure~\ref{f00} shows the numbers of hurricanes making
landfall in each of 39 arbitrary segments along the US coastline during our 54 year period.
If we were to consider smaller segments, or more intense hurricanes only, then the values of $i$ would be smaller.

We will assume that the distribution of the number of hurricanes making landfall is constant in time.
This is probably not true: recent years appear to have seen an increase
in the numbers of hurricanes, for instance.
Our goal, however, is not to deal with that issue (we, and others, have discussed ways to address
that elsewhere), but to look at the basic statistics of this problem.
We will also assume that the number of hurricanes within a year is given by a poisson process.
This is also probably not true, but again making this simplification allows us to look at and think about
some important statistical issues.

To summarise: given $m$ years of historical hurricane data in which there have been
$i$ hurricanes, and assuming stationarity and the poisson distribution, how should we estimate the
probability $f(n)$ of $n$ hurricanes hitting this segment of coastline in a single year? Since
we are assuming that $f$ is a poisson distribution, $f(n)$ is given by:

\begin{equation}\label{f}
f(n)=\frac{e^{-\lambda}\lambda^n}{n!}
\end{equation}

where $\lambda$ is the real unknown expected number of hurricanes per year.

For convenience below we will use the shorthand that:
\begin{equation}
 g(\lambda,n)=\frac{e^{-\lambda}\lambda^n}{n!}
\end{equation}

\section{Classical methods for fitting the poisson distribution}\label{classical}

There are two classical approaches to this problem: method of moments, and maximum likelihood.
We discuss them in turn.

\subsection{Method of moments}

Method of moments works as follows.
We equate the observed annual mean number of hurricanes (which is $\frac{i}{m}$) to the modelled rate $\hat{\lambda}$:
\begin{equation}
 \hat{\lambda}=\frac{i}{m}
\end{equation}
and we substitute this estimate $\hat{\lambda}$ into equation~\ref{f} in place of the real $\lambda$, giving:
\begin{equation}
\hat{f}(n)=\frac{e^{-\hat{\lambda}}\hat{\lambda}^n}{n!}
\end{equation}

That's it: $\hat{f}(n)$ is then our estimate of $f(n)$.

\subsection{Maximum likelihood}

Maximum likelihood works as follows.
The likelihood is defined as the probability of the observed data given the model and the parameters of the model.
Considering likelihood as a function of the parameter $\lambda$, we vary $\lambda$ and find the maximum of the likelihood.
The value of $\lambda$ that maximises the likelihood is then plugged into equation~\ref{f}.

Applying this in practice in our example, we note that we can consider a poisson with rate $\lambda$ for $m$ years
as giving the same distribution of numbers of events as a poisson with rate $\lambda m$, once.

The likelihood is thus given by:
\begin{equation}
L(\lambda)=\frac{e^{-\lambda m} (\lambda m)^i}{i!}
\end{equation}

Taking logs:
\begin{eqnarray}
l&=& log L\\
 &=& log\left[\frac{e^{-\lambda m} (\lambda m)^i}{i!} \right]\\
 &=& -\lambda m + i log (\lambda m) - log (i!)
\end{eqnarray}

Differentiating this by $\lambda$:
\begin{eqnarray}
\frac{\partial l}{\partial \lambda}&=&-m+\frac{i}{\lambda}
\end{eqnarray}

Setting this equal to zero gives:
\begin{equation}
\hat{\lambda}=\frac{i}{m}
\end{equation}

and this is then substituted into equation~\ref{f} in place of the real $\lambda$, giving
\begin{equation}
\hat{f}(n)=\frac{e^{-\hat{\lambda}}\hat{\lambda}^n}{n!}
\end{equation}

\subsection{Discussion of the classical methods}

We see that method of moments and maximum likelihood give the same results.
Do these two methods make sense? Both are based on the idea
that, in order to model the distribution of hurricanes, we should choose our \emph{single
best estimate} of the rate $\lambda$.
Prima facie, this seems reasonable, but it can certainly be criticised.
The main criticism is: why use only our best estimate? What about all the other estimates of the rate
that one might make, which are not quite as likely to be correct? Given the small amount
of data for hurricanes for short sections of coastline,
it seems likely that there is quite a large range of reasonable estimates
for $\lambda$.
Instead of using just the most likely of this range of estimates, perhaps we should use them all in some way.

Another criticism is: consider the case where $i=0$. In other words, we have a section of coastline where
there haven't been any hurricanes in the last $m$ years. What do the classical methods give us?
Our estimate of the rate is $\hat{\lambda}=\frac{i}{m}=\frac{0}{m}=0$, and hence our estimate of the distribution
is $\hat{f}=\frac{e^{-0}0^n}{n!}=0$. In other words, because we haven't seen any hurricanes in the last
$m$ years we conclude that it must be completely impossible that a hurricane might strike us in the future. This is clearly
an illogical conclusion: there are many stretches of coastline that haven't experienced a hurricane strike
in recent times, but for which, on a meteorological basis, it is quite clearly possible that they \emph{could} experience
such a strike.

In fact, this second criticism is related to the first. If there haven't been any hurricanes in the last
$m$ years, it may well be the case that an appropriate conclusion should be that the \emph{most likely} value of
$\lambda$ is $0$,
but it is very clear that we also need to consider the (perhaps small) possibility that $\lambda$ is actually
greater than zero.

One final criticism of the classical methods is: note that the values of $i$ and $m$
only occur as the ratio $\frac{i}{m}$ in these methods.
The absolute value of $m$ doesn't make any difference at all, and so there can be no accounting
for the extra accuracy that extra years of data might bring (or vice versa).

These criticisms lead us onto Bayesian methods, in which we move beyond just considering the single most likely value of $\lambda$,
and consider a range of possible values.

\section{Bayesian methods for fitting the poisson distribution}\label{bayesian}

The Bayesian method we describe works as follows.
Based on the discussion in the previous section, we will express our prediction for the distribution
of the number of hurricanes as an integral over \emph{all possible} hurricane rates that might have given
us the observed data, rather than just the most likely rate.
The prediction is then:

\begin{equation}\label{bayesf}
\hat{f}(n|i) =\int f(n|\lambda) f(\lambda|i) d\lambda.
\end{equation}

The first term in the integral $(f(n|\lambda))$ is the probability of $n$ hurricanes, given a certain value of $\lambda$.
The second term gives the probability of each value of $\lambda$, given the observation of $i$ hurricanes in the
last $m$ years. We can think of this integral as a weighted average of predictions $f(n|\lambda)$,
where $f(\lambda|i)$ gives the weights. $f(\lambda|i)$ is often known as the posterior density
of $\lambda$.

How, then, can we calculate $f(\lambda|i)$?
Applying Bayes' theorem, we can factorise $f(\lambda|i)$ into:

\begin{equation}
 f(\lambda|i) \propto f(i|\lambda)f(\lambda)
\end{equation}

The first term on the right hand side can be evaluated easily: it's just the probability density
of the poisson distribution.

The second term, known as the \emph{prior}, needs a little more thought. If we have prior
information on the distribution of possible values of $\lambda$, we can use this prior
distribution to include that information in the analysis. In this article, however, we will assume that we
have no such prior information. Instead, we will try and choose $f(\lambda)$ so as to be neutral
with respect to different possible values of $\lambda$
(such a choice is often known as a `reference prior', or `uninformative prior').
Unfortunately, there doesn't seem to be
a single unambiguous choice for what this neutral prior should be. We have found the following 3 reasonable suggestions:
\begin{itemize}
    \item $p(\lambda)=c$, a constant value, justified on the basis that this puts equal weights on all possible values of $\lambda$
    \item $p(\lambda)=c {\lambda}^{-\frac{1}{2}}$, known as the Jeffrey's prior, and justified on the basis
    that it is invariant to changes in the scale of $\lambda$
    \item $p(\lambda)=\frac{c}{\lambda}$
\end{itemize}

We can combine these 3 possibilities into one general form:
\begin{equation}
 f(\lambda)=c \lambda^{\alpha}
\end{equation}
where $\alpha$ has the values $0, -1/2$ or $-1$ in our three cases.

This then gives
\begin{eqnarray}
 f(\lambda|i) &\propto  & f(i|\lambda) f(\lambda)\\
              & \propto& c f(i|\lambda) \lambda^{\alpha}
\end{eqnarray}

We can calculate the constant of proportionality $c$ using the fact that the integral of $f(\lambda|i)$ must
be 1 for it to be a probability distribution:
\begin{eqnarray}
 \int f(\lambda|i)d\lambda
  &=&c \int f(i|\lambda) d\lambda\\
  &=&c \int \frac{e^{-\lambda m} (\lambda m)^i \lambda^\alpha}{i!}d\lambda\\
  &=&\frac{c}{m^i i!} \int e^{-\lambda m} \lambda^{i+\alpha}d\lambda\\
  &=&\frac{c}{m^i i!} \int e^{-s} s^{i+\alpha}\frac{ds}{m}\\
  &=& \frac{c m^{-\alpha-1}}{i!} (i+\alpha)!\\
  &=&1
\end{eqnarray}

 implying that
 \begin{equation}
    c=\frac{m^{1+\alpha}i!}{(i+\alpha)!}
\end{equation}

and hence the prior is:
\begin{equation}
f(\lambda)=\frac{m^{1+\alpha}i! \lambda^\alpha}{(i+\alpha)!}
\end{equation}

and the posterior distribution for $\lambda$ is:
\begin{eqnarray}
 f(\lambda|i)&=&\frac{m}{(i+\alpha)!} e^{-\lambda m} (m\lambda)^{i+\alpha}\\
             &=&m g(\lambda m,i+\alpha)
 \end{eqnarray}

Note that this last line should not be read to imply that the distribution of $f(\lambda|i)$ is a poisson distribution,
since now the roles of the parameter and the random variable are switched. In fact this is a gamma distribution.

Now consider the case $\alpha=-1$, $i=0$. The posterior becomes:
\begin{eqnarray}
 f(\lambda|i)&=&\frac{m}{(-1)!} e^{-\lambda m} (m\lambda)^{-1}
 \end{eqnarray}

This is not a proper posterior density, since the integral is not finite, so we reject the $\alpha=-1$
prior, leaving just the cases $\alpha=0$ and $\alpha=-1/2$.

\begin{figure}[!hb]
  \begin{center}
    \scalebox{1.0}{\includegraphics{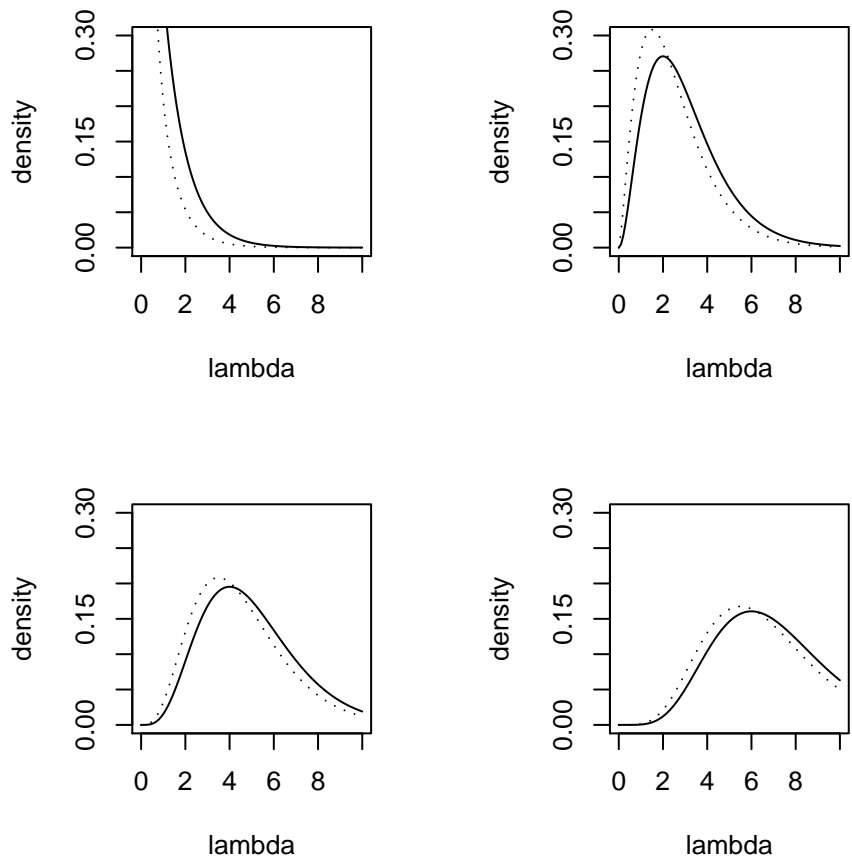}}
  \end{center}
    \caption{
The posterior density for the poisson rate, given 54 years of data
and 0,2,4 and 6 observed hurricanes making landfall over that 54 year period.
     }
     \label{f01}
\end{figure}

Figure~\ref{f01} shows the posterior densities for $m=54$ and $i=0,2,4,6$, for both priors. We see that
the differences are rather small.

Where are the maximum values of these posterior distributions?
Differentiating the prior wrt $\lambda$ gives:
\begin{equation}
 \frac{dp}{d\lambda}=\frac{m^{i-1-\alpha}e^{-\lambda m}\lambda^{i+\alpha-1}}{(i+\alpha)!}[i+\alpha-\lambda m]
\end{equation}
Setting this equal to zero then gives:
\begin{equation}
    \lambda=\frac{i+\alpha}{m}
\end{equation}

For the $\alpha=0$ prior, this gives $\lambda=i/m$, which agrees exactly with the maximum likelihood estimate
of $\lambda$. This is, of course, because the posterior is equal to (or proportional to) the likelihood for a constant prior.

For the $\alpha=-1/2$ prior, this gives the slightly lower value of $\lambda=(i-1/2)/m$. The use of a prior that
weights towards $\lambda=0$ has shifted the maximum slightly relative to the maximum in the likelihood.
The implication is that
this value of $\lambda$ is now the most likely (which we find slightly surprising).

What about the mean value of $\lambda$ under these posterior distributions?

\begin{eqnarray}
\mbox{mean}&=&\int \frac{m}{(i+\alpha)!} e^{-\lambda m} (m\lambda)^{i+\alpha} \lambda d\lambda\\
           &=&\frac{m^{1+\alpha+i}}{(i+\alpha)!} \int \lambda^{i+\alpha+1} e^{-\lambda m} d\lambda\\
           &=&\frac{m^{1+\alpha+i}}{(i+\alpha)!} \int e^{-s} \left(\frac{s}{m}\right)^{i+\alpha+1} \frac{ds}{m}\\
           &=&\frac{1}{m}\frac{(i+\alpha+1)!}{(i+\alpha)!}
\end{eqnarray}

For the $\alpha=0$ prior this gives $\mbox{mean lambda}=(i+1)/m$, while for the $\alpha=-1/2$ prior this gives the
rather unpleasant looking
$\mbox{mean}=\frac{(i+3/2)!}{(i+1/2)!}\frac{1}{m}$.
Note that both of these values are larger than the mean that comes from the classical analysis, which is $i/m$.
This is because the uncertainty wrt the possibility of values of $\lambda$ greater than $i/m$ adds more to the
calculation of the mean than the uncertainty wrt the possibility of values of $\lambda$ lower than $i/m$ takes
away. This is especially noticeable for the case $i=0$, where the two means are both above zero in the Bayesian case.

Substituting the prior into our expression for the Bayesian forecast, equation~\ref{bayesf}, gives:

\begin{eqnarray}
 f(n|i)&=&\int f(n|\lambda)f(\lambda|i)d\lambda\\
       &=&\int g(\lambda,n) m g(\lambda m,i+\alpha) d\lambda\\
       &=&\frac{m^{i+\alpha+1}}{n!(i+\alpha)!}\int e^{-(m+1)\lambda}\lambda^{n+i+\alpha}d\lambda\\
       &=&\frac{m^{i+\alpha+1}}{n!(i+\alpha)!} \left(\frac{1}{m+1}\right)^{i+\alpha+n+1}(i+\alpha+n)!\\
       &=&\frac{(i+\alpha+n)!}{(i+\alpha)!n!}\left(\frac{m}{m+1}\right)^{i+\alpha+1}\left(\frac{1}{m+1}\right)^n\label{nb}
\end{eqnarray}

This is now our estimate for the distribution for the number of hurricanes, based on Bayesian reasoning, and is,
in fact, the negative binomial distribution.

\subsection{Discussion of the Bayesian method}

There are a number of interesting implications of equation~\ref{nb}.

The first is simply that, to predict, or `model' the poisson distribution, one shouldn't actually use the poisson
distribution: one should use the negative binomial instead.\footnote{Although it is important to realise that
the negative binomial in this case is fitted using $i$ and $m$, and not using the standard methods for fitting
the negative binomial such as method of moments, which would give different parameter values and would only be valid
if we thought that the negative binomial, not the poisson, was a good model for the underlying data.}.
This is because of parameter uncertainty. The fact
that we don't know the exact value of the parameter $\lambda$, and that we represent that uncertainty using
a distribution, converts the poisson into the negative binomial (even if we are still considering the events
to be independent).

The second interesting implication, already mentioned above, is that the mean of our forecast is now higher
than for the classical case. The mean of the classical forecast (which is $\frac{i}{m}$)
arises simply because the mean is given by the value of the rate, and the value of the rate is chosen to
be the maximum likelihood value of $\frac{i}{m}$. The Bayesian mean, on the other hand, arises through consideration
of exactly what can be inferred from the observation of $i$ hurricanes in $m$ years. We feel that the Bayesian
forecast mean is thus much more strongly justified than the classical mean.

One particular case of interest is when $i=0$: in this case the mean predicted number of hurricanes is
created entirely from the possibility of the rate being higher than the most likely estimate of the rate.

The fourth interesting implication is the variance of the forecast.

The variance can be calculated as:
\begin{eqnarray}
\mbox{variance}&=&\int \frac{m}{(i+\alpha)!} e^{-\lambda m} (m\lambda)^{i+\alpha} \lambda^2 d\lambda\\
               &=&\frac{m^{1+\alpha+i}}{(i+\alpha)!} \int \lambda^{i+\alpha+2} e^{-\lambda m} d\lambda\\
               &=&\frac{m^{1+\alpha+i}}{(i+\alpha)!} \int e^{-s} \left(\frac{s}{m}\right)^{i+\alpha+2} \frac{ds}{m}\\
               &=&\frac{1}{m^2}\frac{(i+\alpha+2)!}{(i+\alpha)!}
\end{eqnarray}

This is larger than the variance of the classically fitted poisson, which is $\frac{i^2}{m^2}$, and
the ratio of the variance to the mean is not 1, as
it is for the poisson, but is greater than 1.

The fifth interesting implication is that we now have a non-trivial distribution for the possible number
of future hurricanes, even in the case when $i=0$, given by:

\begin{eqnarray}
 p(n|i=0)&=&\frac{(i+\alpha+n)!}{(i+\alpha)!n!}\left(\frac{m}{m+1}\right)^{i+\alpha+1}\left(\frac{1}{m+1}\right)^n\\
         &=&\frac{(\alpha+n)!}{(\alpha)!n!}\left(\frac{m}{m+1}\right)^{\alpha+1}\left(\frac{1}{m+1}\right)^n
\end{eqnarray}

So, interestingly, even if there have never been any hurricanes in the past, this method still predicts
a non-zero probability for hurricanes in the future.
This seems reasonable: since $m$ is finite, we can't rule out the possibility that we haven't just been
very lucky over the last $m$ years and avoided any hurricane strikes.

\section{Numerical comparison}\label{numerical}

\begin{figure}[!hb]
  \begin{center}
    \scalebox{1.0}{\includegraphics{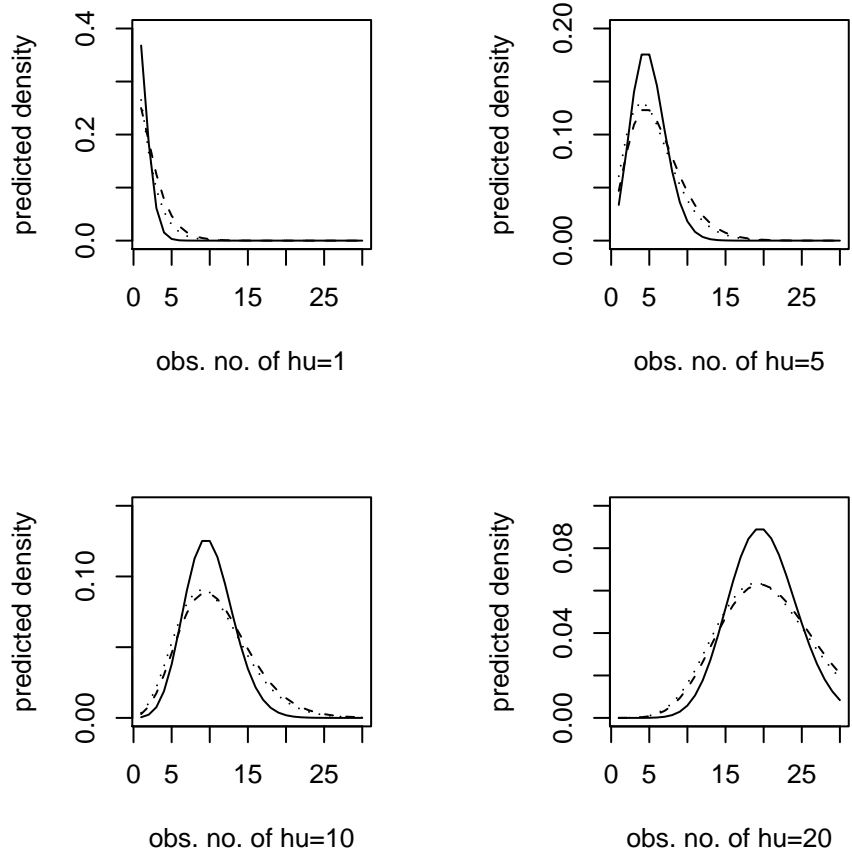}}
  \end{center}
    \caption{
Predictions of future hurricane numbers, based on either 1, 5, 10 or 20
historical hurricanes making landfall in a 54 year period.
The solid line shows classical predictions and
the dotted line shows Bayesian predictions.
     }
     \label{f02}
\end{figure}

In this section, we perform a numerical comparison of forecasts from the classical and Bayesian methods, for different
values of $i$. We ask the question: how much difference does it really make to the final probabilities we predict
if we use the Bayesian method? We have already seen that for $i=0$ there is a big difference between the two
methods (the classical method predicts zero probabilities for $n>0$, while the Bayesian methods predict non-zero probabilities)
but what about for larger values of $i$?.
Figure~\ref{f02} shows predictive distributions from the classical and Bayesian methods, for $m=54$, and values
of $i$ of 1,5, 10 and 20. We see large differences between the classical and Bayesian methods for all the values of $i$
tested. The Bayesian methods give a flatter, broader distribution, as expected, and the probability of extreme
numbers of hurricanes is much higher.
The differences between the two Bayesian methods, however, are much smaller, with the $\alpha=0$ method giving
a slightly flatter, broader distribution.

At this point, we weigh up the pros and cons of the two priors that we are still considering, and
make a decision as to which to use. In favour of the $\alpha=-1/2$ prior, there is a piece of (slightly esoteric)
theoretical reasoning related to the Fisher information and transformations of scale.
In favour of the $\alpha=0$, we have that:

\begin{itemize}
    \item a flat prior in $\lambda$ reflects our desire to avoid
    having the prior influence the final result, while the $\alpha=-1/2$ prior weights
    towards lower values of $\lambda$
    \item the flat prior gives an intuitively reasonable value of the most likely value
    of $\lambda$, while the $\alpha=-1/2$ gives a most likely value that conflicts with intuition
    \item the mathematics is simpler
    \item the differences between forecasts made by the two priors is very small (relative to the difference
    between classical and Bayesian priors)
    \item it is slightly more conservative (i.e. gives wider distributions)
\end{itemize}

Based on this we conclude that we prefer the $\alpha=0$ prior.

For convenience, we now summarise the key properties of this prior.

\subsection{Summary of properties for the uniform prior}

\begin{eqnarray}
 \mbox{prior}&=&m\\
 \mbox{posterior}&=&\frac{m}{i!} e^{-\lambda m} (m\lambda)^i\\
 \mbox{predictive probability}&=&\frac{(i+n)!}{(i+)!n!}\left(\frac{m}{m+1}\right)^{i+1}\left(\frac{1}{m+1}\right)^n\\
 \mbox{forecast mean}&=&\frac{i+1}{m}\\
 \mbox{forecast variance}&=&\frac{(i+1)(i+2)}{m^2}\\
 \mbox{maximum of the posterior}&=&\frac{i}{m}
\end{eqnarray}

\section{Scoring probabilistic forecasts}\label{scoring}

We now change tack slightly, and consider how one might evaluate
a prediction of the distribution of the number of hurricanes.
The scoring system we will use is based on the out-of-sample log-likelihood, which we have
previously discussed and used in a number studies, such as~\citet{j94}. We believe this is the most
obvious score to use for comparing probabilistic forecasts, and it is closely related
to a large body theory concerning the \emph{cross-entropy} and the \emph{Kullback-Leibler divergence}.

The out-of-sample likelihood is defined loosely as the expectation of the log of the probability
of the observations given the forecast. This is only a loose definition because to make it precise
we need to specify what we mean by expectation, and there are several possibilities.
This is discussed in more detail in~\citet{j94}.
In this section we will define the expectation as being over all future values, holding the historical data (the value
of $i$) fixed, and over all possible values for the unknown parameters (this is the score $S_3$ in~\citet{j94}).

%
%
%
%
%

For the classical prediction methods, the score is:
\begin{eqnarray}
 S_3
    &=&\int mg(\lambda m,i) \left( \sum_{n} g(\lambda,n) \mbox{ log } g
    \left(\frac{i}{m},n\right) \right) d\lambda\\
    &=&
       m\sum_n \left( \int g(\lambda m,i)
       g(\lambda,n) \mbox{ log } g
       \left(\frac{i}{m},n\right) d\lambda \right)\\
    &=&
       m\sum_n \mbox{ log } g
       \left( \frac{i}{m},n\right)
       \int g(\lambda m,i)
       g(\lambda,n) d\lambda \\
    &=&
       m\sum_n \mbox{ log } g
       \left( \frac{i}{m},n\right)
       \int \frac{e^{-\lambda m} (\lambda m)^i}{i!}
                   \frac{e^{-\lambda} \lambda^n}{n!} d\lambda
       \\
    &=&
       \sum_n \mbox{ log } g
       \left( \frac{i}{m},n\right)
       \frac{m^{i+1}}{i!n!} \int e^{-\lambda (m+1)} \lambda^{i+n} d\lambda
       \\
    &=&
       \sum_n \mbox{ log } g
       \left( \frac{i}{m},n\right)
       \frac{m^{i+1}}{i!n!} \left(\frac{1}{m+1}\right)^{i+n+1} \int e^{-\lambda} \lambda^{i+n} d\lambda
       \\
    &=&
       \sum_n \mbox{ log } g
       \left( \frac{i}{m},n\right)
       \frac{m^{i+1}}{i!n!} \left(\frac{1}{m+1}\right)^{i+n+1} (i+n)!
       \\
    &=&
       \sum_n \mbox{ log } g
       \left( \frac{i}{m},n\right)
       \frac{(i+n)!}{i!n!} \left(\frac{m}{m+1}\right)^{i+1} \left(\frac{1}{m+1}\right)^n
\end{eqnarray}

For the Bayesian prediction method the score is:

\begin{eqnarray}
 S_3
    &=&\int m g(\lambda m,i)
       \left( \sum_n g(\lambda,n) \mbox{log}
       \left[\frac{(i+n)!}{i!n!} \left(\frac{m}{m+1}\right)^{i+1}\left(\frac{1}{m+1}\right)^n
       \right]
       \right)\\
    &=&\int m \frac{e^{-\lambda m}(\lambda m)^i}{i!}
       \left( \sum_n \frac{e^{-\lambda}\lambda^n}{n!} \mbox{log}
       \left[\frac{(i+n)!}{i!n!} \left(\frac{m}{m+1}\right)^{i+1}\left(\frac{1}{m+1}\right)^n
       \right]
       \right) d\lambda\\
    &=&\frac{m^{i+1}}{i!}
       \sum_n \frac{1}{n!} \mbox{log}
       \left[
       \frac{(i+n)!}{i!n!} \left(\frac{m}{m+1}\right)^{i+1} \left(\frac{1}{m+1}\right)^n
       \right]
       \int e^{-(m+1)\lambda}\lambda^{n+i}d\lambda\\
    &=&\frac{m^{i+1}}{i!}
       \sum_n \frac{1}{n!} \mbox{log}
       \left[
       \frac{(i+n)!}{i!n!} \left(\frac{m}{m+1}\right)^{i+1} \left(\frac{1}{m+1}\right)^n
       \right]
       \left(\frac{1}{m+1}\right)^{i+n+1}(i+n)!\\
    &=&\left(\frac{m}{m+1}\right)^{i+1}
       \sum_n \left(\frac{1}{m+1}\right)^{n} \frac{(i+n)!}{i!n!} \mbox{log}
       \left[
       \frac{(i+n)!}{i!n!} \left(\frac{m}{m+1}\right)^{i+1} \left(\frac{1}{m+1}\right)^n
       \right]
\end{eqnarray}

\begin{figure}[!hb]
  \begin{center}
    \scalebox{0.7}{\includegraphics{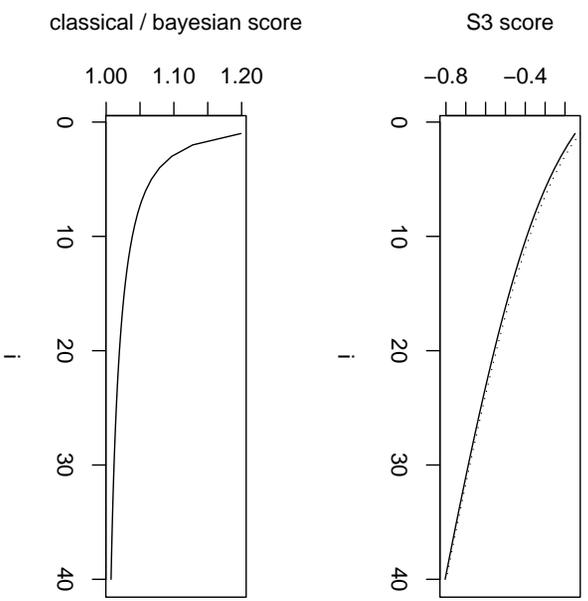}}
  \end{center}
    \caption{
Theoretical scores from the Bayesian and classical predictions.}
     \label{f033}
\end{figure}

We now plot these scores (the classical forecast score and the Bayesian forecast score)
for $m=54$ and for a few values of $i$, in figure~\ref{f033}. We see that the Bayesian score always beats the
classical score, but the difference between the scores reduces as $i$ increases.

\section{Empirical comparison}\label{empirical}

Finally, we compare the classical and Bayes prediction methods for real data.
Our comparison is based on a carefully constructed cross-validation procedure (the Quenouille-Tukey jackknife)
to make it as fair and realistic as possible. It works as follows:
\begin{itemize}
    \item we loop over the 54 years of data, missing out each year in turn
    \item we fit the two models to the remaining 53 years of data
    \item we calculate the log of the probability of the 54th year of data for both models
    \item we average together the logs of the probabilities over the 54 years
    \item we repeat this for each of our 39 gates along the US coastline
\end{itemize}

The scoring system being used can be considered as an empirical version of the predictive log-likelihood score that
we use in section~\ref{scoring}, where the expectation is now over the loop in the jackknife
(which doesn't, however, correspond exactly to the definition of expectation used in section~\ref{scoring}).

\begin{figure}[!hb]
  \begin{center}
    \scalebox{0.7}{\includegraphics{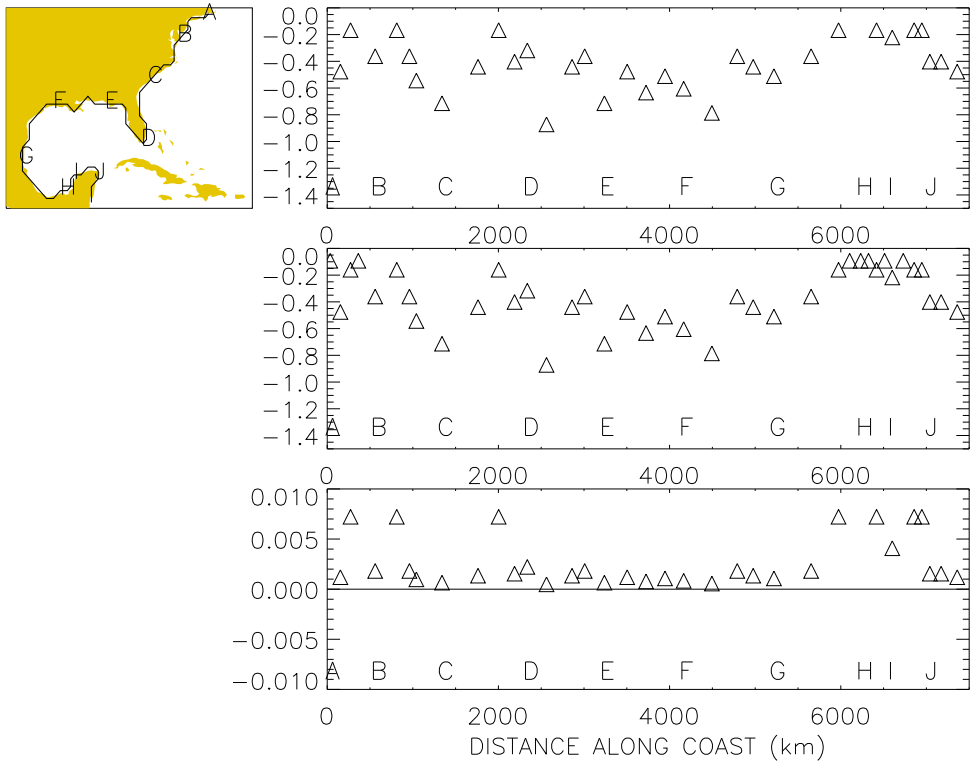}}
  \end{center}
    \caption{
Performance of the classical and Bayesian prediction methods, evaluated using
cross-validation, for the number of hurricanes crossing our 39 coastline
segments. The score is the out-of-sample expected
predictive log-likelihood. The top right panel shows the score for the classical
prediction, the middle right panel shows the score for the Bayesian prediction, and
the bottom right panel shows the difference between the two.
We see that the Bayesian method wins for every segment.
     }
     \label{f03}
\end{figure}

Results from this comparison are shown in figure~\ref{f03}. We see that the Bayesian method beats the classical
method for each one of the 39 gates. The difference are largest for the segments where there are fewest
hurricanes, as we'd expect.

\begin{figure}[!hb]
  \begin{center}
    \scalebox{0.7}{\includegraphics{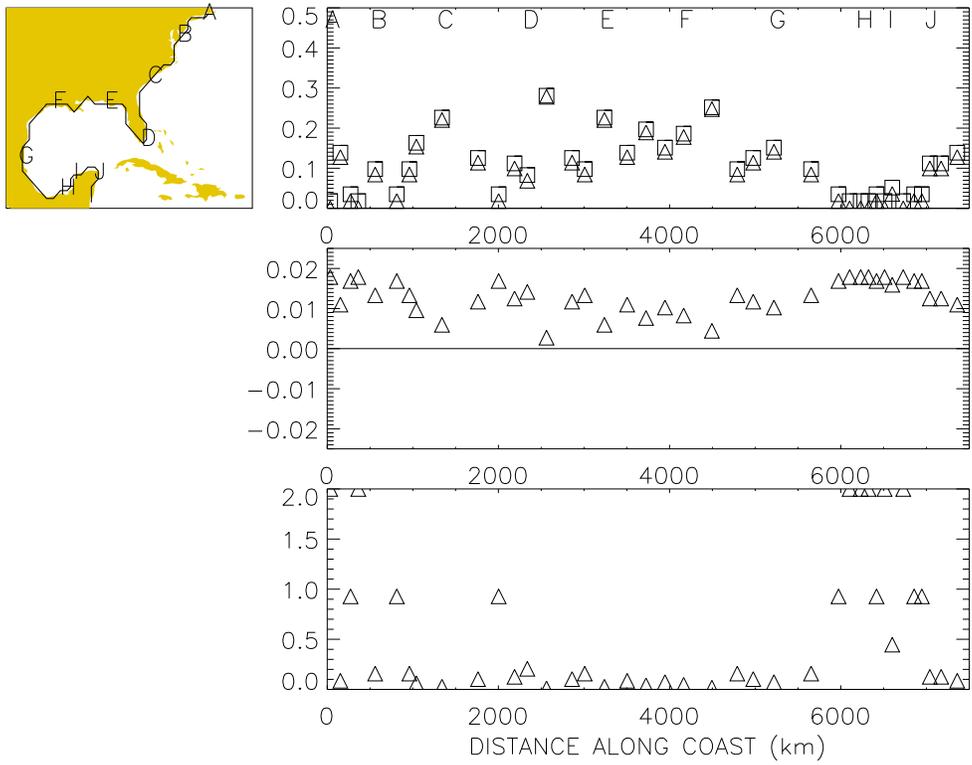}}
  \end{center}
    \caption{
Probabilities of one hurricane making landfall in a year, from classical and Bayesian predictions.
The top panel shows the probabilities themselves (classical=cirles, Bayesian=squares).
The middle panel shows the differences, and the bottom panel shows the differences
divided by the classical probabilities. In the bottom panel values of 2 are actually infinity.
We see that the Bayesian probabilities are
all higher, although only by a small absolute amount. The relative differences are also small
for most gates, except where there are either no historical hurricanes, or a very small number of
historical hurricanes.
     }
     \label{f04}
\end{figure}

Finally, figure~\ref{f04} shows the actual probabilities predicted for the occurrence of 1 hurricane by gate for the
classical and Bayesian methods. We see that the probabilities are mostly reasonable similar, but that big differences
occur where there are very few historical hurricanes.

\section{Discussion}\label{discuss}

We have discussed the question of how to predict the distribution of the number of hurricanes crossing
segments of the US coastline. We have taken a very simple approach based on the assumptions that events
are independent, and that the underlying poisson rates are constant in time. This simple framework allows us to
compare classical and Bayesian statistical methods in some detail. We derive expressions for the classical
and Bayesian forecasts, and further expressions for their performance under an expected predictive log-likelihood
scoring system. At a theoretical level we find that the Bayesian method performs better.
We then test our classical and Bayesian
prediction methods on real hurricane data, using a jack-knife cross-validation scheme. We find that the Bayesian
method does better in practice too, but that again the differences in the forecasts are only significant for those sections of
coastline with very low historical hurricane rates.

The next stage of our research is to compare these forecasts of landfall rates with forecasts derived from
the basin-wide track model we have described in a series of recent papers~\citep{hallj05f}.

\bibliography{arxiv}

\newpage

\newpage

\newpage

\newpage

\newpage

\newpage

\end{document}